\documentclass[aps,prl,reprint,amsmath,amssymb,superscriptaddress,noeprint,nofootinbib,floatfix]{revtex4-2}
\def\bibsection{\section*{\refname}} 
\pdfoutput=1

\usepackage{graphicx}
\usepackage{dcolumn}
\usepackage{bm}
\usepackage[version=3]{mhchem}
\usepackage{mathtools}
\PassOptionsToPackage{hyphens}{url}
\usepackage[normalem]{ulem}
\usepackage{silence}
\usepackage{float}
\usepackage{color}
\usepackage{amsmath,amssymb}
\usepackage{upgreek}
\usepackage{psfrag} 
\usepackage{graphicx}
\usepackage{braket}
\usepackage{units}
\usepackage{times}
\usepackage{tabularx}
\usepackage{upgreek}
\usepackage[svgnames]{xcolor}

\definecolor{myColor}{rgb}{0.02,0.12,0.3}
\definecolor{myciteColor}{rgb}{0.39,0.7,0.89}
\usepackage[colorlinks=true,allcolors=blue]{hyperref}

\newcommand{\dd }{\mathrm{d}}

\newcommand{\as}{a_\mathrm{s}}

\newcommand{\Nb}{N_\mathrm{b}}
\newcommand{\Ei}{E_\mathrm{i}}

\newcommand{\be}{\begin{equation}}
\newcommand{\ee}{\end{equation}}
\newcommand{\bea}{\begin{align}}
\newcommand{\eea}{\end{align}}

\usepackage{lineno}

\makeatletter
\def\maketitle{
\@author@finish
\title@column\titleblock@produce
\suppressfloats[t]}
\begin{document}

\title{Core-bound waves on a Gross-Pitaevskii vortex}

\author{Evan Papoutsis}
\thanks{These authors contributed equally to this work}
\affiliation{
Department of Physics, Yale University, New Haven, Connecticut 06520, USA}

\author{Nathan Apfel}
\thanks{These authors contributed equally to this work}

\affiliation{
Department of Physics, Yale University, New Haven, Connecticut 06520, USA}

\author{Nir~Navon}

\affiliation{
Department of Physics, Yale University, New Haven, Connecticut 06520, USA}
\affiliation{Yale Quantum Institute, Yale University, New Haven, Connecticut 06520, USA}

\date{\today}

\begin{abstract} 

We find the dispersion relations of two elusive families of core-bound excitations of the Gross-Pitaevskii (GP) vortex, $\text{\textit{varicose}}$ (axisymmetric) and \textit{fluting} (quadrupole) waves. For wavelengths of order the healing length, these two families---and the well-known Kelvin wave---possess an infinite sequence of core-bound, vortex-specific branches whose energies lie below the Bogoliubov dispersion relation. In the short-wavelength limit, these excitations can be interpreted as particles radially bound to the vortex, which acts as a waveguide. In the long-wavelength limit, the fluting waves unbind from the core, the varicose waves reduce to phonons propagating along the vortex, and the fundamental Kelvin wave is the only  core-bound vortex-specific excitation. Finally, we propose a realistic spectroscopic protocol for creating and detecting the varicose wave, which we test by direct numerical simulations of the GP equation. 

\end{abstract}
\maketitle

Quantum vortices---topological defects that carry quantized circulation---play a central role in the dynamics of superfluids and superconductors. Proposed by Onsager in 1949, they govern how a superfluid responds to rotation and how it dissipates energy. Importantly, vortices are dynamic objects, \textit{i.e.} they support excitations. The archetypal example is the Kelvin wave, a helical deformation of a vortex line. Kelvin waves have been studied extensively in theory~\cite{Pitaevskii1961} and in experiment~\cite{bretin2003quadrupole,fonda2014direct,minowa2025direct}, and are widely believed to play an important role in the dissipation of quantum turbulence~\cite{vinen2002quantum,Barenghi2023}.

The properties of Kelvin waves are generically insensitive at long wavelengths to the structure of the vortex core. However, this structure can give rise to interesting phenomena---such as bound states~\cite{caroli1964bound}---that can feed back on vortex dynamics. Yet, even within the simplest model of superfluidity---the Gross-Pitaevskii (GP) equation---the nature of structure-dependent excitations remains unsettled.

It has been speculated that a GP vortex may also support an axisymmetric excitation: an oscillation of the core radius called a \emph{varicose wave}. For classical vortices, Lord Kelvin solved the corresponding problem and showed that the existence and behavior of such waves depend sensitively on the core structure~\cite{Thompson1880}. For quantum vortices, however, the status of varicose waves has been marred by confusion. A variational treatment suggested that GP vortices support varicose waves~\cite{fetter1965vortices}, whereas a classic textbook later questioned their existence~\cite{Donnelly1991}. A linear analysis of the GP equation in a cylindrical trap identified Kelvin waves but no varicose excitations~\cite{Isoshima1997}. Subsequent simulations reported varicose-like excitations in a harmonically trapped Bose–Einstein condensate (BEC). However, they were recognized as modes related to the external trapping potential: they were not vortex-specific~\cite{Simula2008}. A third kind of excitation---a quadrupole fluting wave---has been suggested but not studied~\cite{fetter1965vortices}.

\begin{figure}[!hbt]
\includegraphics[width=1.02\columnwidth]{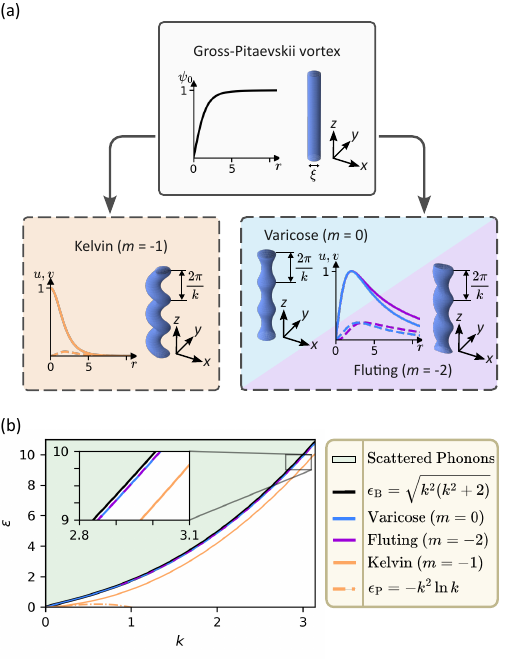}
\caption{\textbf{Excitations of a Gross-Pitaevskii vortex} (a) (top) The wavefunction of the condensate with the vortex $\psi_0$ and a 3D density contour. (bottom) Radial excitation wavefunctions $u$ (solid) and $v$ (dashed)---see text---with the density contours for the corresponding excited vortex. (b) The dispersion relations of the fundamental branches of these excitations $\epsilon^{(n=0)}_{m}(k)$. We show the Bogoliubov dispersion relation $\epsilon_{\mathrm B} = \sqrt{k^2(k^2+2)}$ (black line), the scattered phonons (green region), the varicose branch $m=0$ (blue line), and the fluting branch $m=-2$ (purple line). The Kelvin branch $m=-1$ is shown as an orange line and its long-wavelength approximation $\epsilon_{\mathrm P} = -k^2\ln{k}$ as the dot-dashed orange line. Inset: zoomed-in dispersion relations.}
\label{Fig1}
\end{figure}

In this Letter, we show that core-bound varicose and fluting excitations exist on a GP vortex at sufficiently small wavelengths. 
We analyze the corresponding eigenvalue problem for both infinite- and finite-size systems and calculate their dispersion relations. We confirm these predictions by direct simulations of the GP equation.

Our starting point is the dimensionless GP equation
\begin{equation}
    i\partial_t \Psi(\mathbf r,t) = \big(-\nabla^2+|\Psi(\mathbf r, t)|^2\big)\Psi(\mathbf r,t),
    \label{GPE}
\end{equation}
which describes the mean-field wavefunction $\Psi$ of a weakly interacting BEC; here, we set the chemical potential of the condensate $\mu$ and the healing length $\xi = \hbar/\sqrt{2M\mu}$ to $1$ (where $M$ is the boson mass) and normalize $\Psi$ such that $\int |\Psi|^2\mathrm{d}V = gN/\mu$, where $g=4\pi\hbar^2 \as/M$ is the coupling constant, $\as$ is the s-wave scattering length, and $N$ the total particle number (see Supplemental Material~\cite{supp} for details).

We linearize Eq.~(\ref{GPE}) around a singly-quantized straight vortex line at $r=0$: $\Psi(r,\varphi,z,t) = e^{i(\varphi - t)}(\psi_0(r)+\delta \psi(r,\varphi,z,t))$, where $\psi_0$ is the vortex wavefunction [Fig.~\ref{Fig1} (a)]. The perturbation $\delta \psi$ can be written as
\begin{eqnarray}
    \delta \psi(r,\varphi,z,t)&=& u^{(n)}_m(k,r)e^{i(m\varphi +kz - \epsilon^{(n)}_m(k)t)}\nonumber\\
    &&-v^{(n)*}_m(k,r)e^{-i(m\varphi+kz-\epsilon^{(n)}_m(k)t)},
\end{eqnarray}
where $k$ and $m$ are the wavenumbers along $z$ and $\varphi$, $n$ is the branch index, and $\epsilon^{(n)}_m(k)$ is the energy. The excitation amplitudes $u^{(n)}_m(k,r)$ and $v^{(n)}_m(k,r)$ are obtained by solving the equation
\begin{equation}
    \begin{pmatrix}
        \mathcal L_{m}&-\psi_0^2 \\-\psi_0^2&\mathcal L_{-m}
    \end{pmatrix}\begin{pmatrix}
        u\\v
    \end{pmatrix} = \epsilon\begin{pmatrix}
         u\\-v
    \end{pmatrix}, 
    \label{u,v}
\end{equation}
where $\mathcal L_m \equiv -\frac{1}{r}\partial_r(r\partial_r) + \frac{(1+m)^2}{r^2}+k^2+2\psi_0^2(r)-1$; here, we dropped  $m$, $n$ and $k$ from $u$, $v$ and $\epsilon$ for notational clarity.

Our search for core-bound vortex excitations is guided by two qualitative criteria. First, they should be localized, \textit{i.e.} $u(r)$ and $v(r)$ should decay exponentially at large $r$. Second, they should be vortex-specific. In an infinite system---for which $\psi_0(r\to\infty)=1$---these requirements reduce to a single inequality: $\epsilon(k)<\epsilon_{\mathrm B}(k)\equiv \sqrt{k^2(k^2+2)}$, where $\epsilon_{\mathrm B}$ is the Bogoliubov dispersion relation. For $\epsilon<\epsilon_{\mathrm B}$, solutions to Eq.~(\ref{u,v}) behave as a modified Bessel function $K$, which decays for large $r$. For $\epsilon>\epsilon_{\mathrm B}$, physically relevant solutions oscillate like unmodified Bessel functions $J$ and $Y$, forming a continuum of scattering states. Moreover, in the absence of a vortex there are no excitations below $\epsilon_{\mathrm B}$, so the inequality also enforces vortex specificity. For the GP model, localized vortex-specific excitations come in three flavors: $m=0$ (varicose), $m=-1$ (Kelvin), and $m=-2$ (fluting). Their radial profiles are sketched in Fig.~\ref{Fig1}(a). 

In Fig.~\ref{Fig1}(b), we show our main result: the dispersion relations of these excitations. We solved the eigenvalue problem Eq.~(\ref{u,v}) with a pseudo-spectral method in a Laguerre polynomial basis, for which the kinetic term is evaluated analytically, and the potential term is computed using a Gaussian quadrature rule~\cite{Almarzoug2011,Alhaidari2008} (see Supplemental Material~\cite{supp}). The blue line is the lowest $m=0$ mode, the axisymmetric varicose wave. Since its energy $\epsilon(k)$ is smaller than $\epsilon_{\mathrm B}(k)$ for all $k>0$, it is core-bound. The purple line is the $m=-2$ fluting mode, which is close to the varicose dispersion relation at large $k$. The Kelvin wave ($m=-1)$ is shown in orange. It approaches Pitaevskii's analytic result $\epsilon_\mathrm{P}(k)$ at small $k$ (dot-dashed orange)~\cite{Pitaevskii1961} and becomes quadratic for large $k$.

\begin{figure}[!hbt]
\includegraphics[width=\columnwidth]{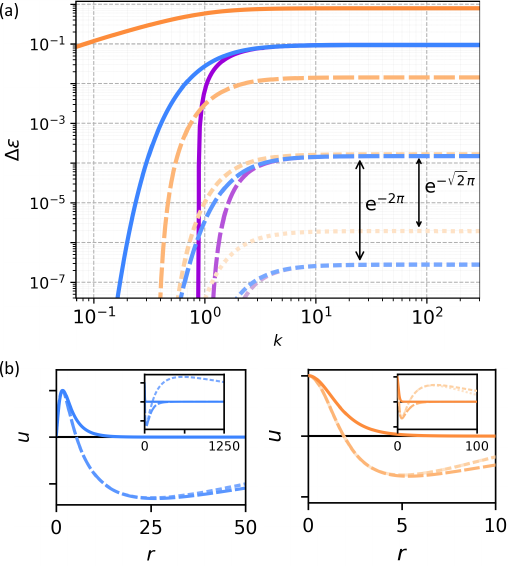}
\caption{\textbf{Spectrum of the GP vortex excitations} (a) Binding energies $\Delta \epsilon \equiv \epsilon_{\mathrm B}-\epsilon$ of varicose waves (blue; $m=0$), fluting waves (purple; $m=-2$) and Kelvin waves (orange; $m=-1$); the fundamental branches are shown as solid lines and the higher branches are dashed, with shorter dashes indicating higher $n$. The two double arrows show ratios of $\mathrm{e}^{-2\pi}$ and $\mathrm{e}^{-\sqrt{2}\pi}$ (see text).  Note that despite appearances, in the limit $k\rightarrow\infty$, $\Delta\epsilon_{-1}^{(2)}\neq\Delta\epsilon_0^{(1)}$. (b) The spatial profiles $u(r)$ of the first three $m=0$ (resp. $m=-1$) modes in the limit $k\rightarrow \infty$ are shown on the left (resp. on the right); in this limit $v \rightarrow 0$. The color code and line styling follow the same convention as in (a). Insets: Zoomed-out views of the same plots. The normalization is arbitrary.}
\label{Fig2}
\end{figure}

To gain more insight into the structure of the core-bound states, we focus on the binding energy $\Delta \epsilon^{(n)}_m(k) \equiv \epsilon_{\mathrm B}(k)-\epsilon^{(n)}_m(k)$, shown in Fig.~\ref{Fig2}(a). The fundamental branches for each mode ($n=0$) are plotted as solid lines, and the higher branches are shown as long-dashed ($n=1$), short-dashed ($n=2$), and dotted ($n=3$) lines; the color code is the same as in Fig.~\ref{Fig1}(b).

This plot exhibits notable features. First, at large $k$, the binding energy for each branch tends to a finite limit. The largest is the fundamental Kelvin wave's $\Delta\epsilon^{(0)}_{-1}\approx 0.79$, followed by the fundamental varicose's and fluting waves' $\Delta\epsilon^{(0)}_0= \Delta\epsilon^{(0)}_{-2} \approx 0.094$. 
Second, the ratios between successive Kelvin branches are similar ($\Delta\epsilon^{(n+1)}_{-1}/\Delta\epsilon^{(n)}_{-1}\approx e^{-\sqrt{2}\pi}$) suggesting a geometric spacing; the corresponding ratio for the varicose and fluting modes appears to be $e^{-2\pi}$. Third, at small $k$, the fundamental varicose smoothly decreases whereas the fundamental fluting branch's $\Delta\epsilon$ abruptly drops to zero at $k\approx 1$, indicating delocalization. All three features can be understood by analyzing Eq.~\eqref{u,v} in appropriate limits.

First, in the $k\rightarrow \infty$ limit, $v=0$, and Eq.~\eqref{u,v} reduces to a Schrödinger-like equation for $u$: 
\begin{equation}
    \left(-\frac{1}{r}\partial_r(r\partial_r)+U_{\mathrm{eff}}(r)\right)u=-\Delta\epsilon \,u,\label{Eq4}
\end{equation}
with an effective potential $U_{\mathrm{eff}}(r)=2\psi_0^2(r)+(1+m)^2/r^2-2$. This form provides a compelling physical picture: the vortex-induced density depletion generates an attractive potential that can bind modes to the core. It also clarifies why Kelvin modes are more deeply bound: for $m=-1$, the centrifugal barrier $(1+m)^2/r^2$ vanishes. Conversely, no bound states occur for other azimuthal wavenumbers, since the effective potential is repulsive when $m \notin\{ 0,-1,-2\}$.

Second, since $\psi^2_0(r)\sim 1-1/r^2$ at large $r$, the effective potential has an inverse-square tail: $U_\mathrm{eff}(r) \sim -(1-2m-m^2)/r^2$. This has momentous consequences: the spectra of potentials that behave as $1/r^2$ at long distances have universal features independent of short-range details~\cite{case1950singular}, as exhibited for instance in the famous Efimov effect~\cite{EFIMOV1970}. When this potential is attractive---here $m \in\{ 0,-1,-2\}$---one obtains an infinite ladder of bound states with asymptotically geometric spacing. The ratios between adjacent energy levels approach $e^{-2\pi/\sqrt{1-m^2-2m}}$, explaining the spacings in Fig.~\ref{Fig2}(a). The radial profiles of the bound states also grow geometrically, as seen in Fig.~\ref{Fig2}(b). They exhibit log-oscillations up to a radial turning point whose location scales by a factor $e^{\pi/\sqrt{1-m^2-2m}}$ between successive modes.

So far, we set $v=0$, which is valid only in the large-$k$ limit.  A more general approximation that works for all $k$ is $v(r)\sim u(r)(\sqrt{\epsilon^2+1}+\epsilon)^{-1}\left(1-r^{-2}(\epsilon-2m)/\sqrt{\epsilon^2+1}\right)$ at large $r$ (see Appendix A in End  Matter). It is therefore convenient to treat $\epsilon$ as a parameter and solve Eq.~\eqref{u,v} for $k$. 
In the large-$r$ limit, the resulting equation has the same form as Eq.~(\ref{Eq4}) with $\Delta \epsilon$ replaced by $\Delta k^2\equiv k^2+1-\sqrt{\epsilon^2+1}$, and with an effective potential $U_{\mathrm{eff}}(r)\sim -\alpha_m^2(\epsilon)/r^2$, where $\alpha_m^2(\epsilon)=1-m^2-(1+2m\epsilon)/\sqrt{\epsilon^2+1}$. 
When $\alpha_m^2(\epsilon)>0$, the long-range attraction again produces an infinite sequence of bound states, and $\Delta k^2$ scales geometrically with ratio $e^{-2\pi/\alpha_m(\epsilon)}$.   
The short-range behavior of $u$ and $v$ is generally complicated (and depends on $\Delta k^2$), but this does not spoil the geometric structure guaranteed by the attractive long-range tail. When $\alpha_m^2(\epsilon)<0$, by contrast, the existence of bound states is contingent on short-range physics.

Importantly, $\alpha_0^2(\epsilon)>0$ for all nonzero $\epsilon$ but tends to zero as $\epsilon\to 0$. This explains our third observation: varicose modes exist for all nonzero $k$, but their binding energies decrease with $k$.  In the limit $k\rightarrow 0$, all varicose branches approach $u=v=\psi_0$---the Goldstone mode of the U(1) symmetry broken by the vortex wavefunction~\cite{Takahashi2014,Kobayashi2014,Takahashi2015,sym} (in a finite system, \emph{i.e.} setting $\Psi(r=R)=0$, each varicose dispersion relation is gapped and crosses $\epsilon_\mathrm{B}(k)$ at a finite $k$ that goes to zero in the infinite-system limit, see End Matter).  By contrast, we find core-bound fluting modes only for $k\gtrsim 1$. For $k \lesssim 1$, their energy lies above $\epsilon_\mathrm{B}$: the long-range $m=-2$ potential is repulsive $(\alpha^2_{-2}(\epsilon)<0)$ and the short-range part is presumably not deep enough to hold them.

\begin{figure}[hbt!]
\includegraphics[width=\columnwidth]{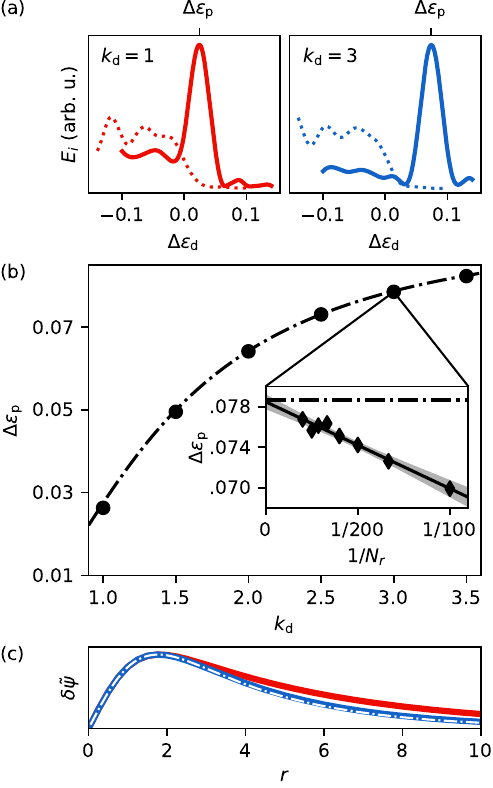}
\caption{\textbf{Numerical spectroscopy of varicose waves.} (a) Typical response curves for the injected energy $E_\mathrm{i}$ versus drive frequency $\Delta\epsilon_\mathrm{d}$ are plotted for $k_\mathrm{d}=1$ (red) and $k_\mathrm{d}=3$ (blue), with (solid) and without (dashed) a vortex. Each spectrum is driven with $r_\mathrm{d}=0.25$ for $T=150$ (the Fourier-limited FWHM is $\approx 0.04$); here $N_r=200$, $N_z=12$, and $N_t=2\times10^5$. (b) The spectroscopically extracted dispersion relation is shown as solid circles; the linear analysis calculation is the dash-dotted line. Inset: the varicose peak position versus radial grid points $1/N_r$ for $k_\mathrm{d}=3$, together with a linear fit for the $N_r\to\infty$ extrapolation. Each resonance was driven with $r_\mathrm{d}=5$ for $T=100$ (the Fourier-limited FWHM is $\approx 0.06$). Throughout this panel, $N_z=12$ and $N_t=3.5\times 10^5$. The gray band is the $95\%$ confidence interval on the fitted parameters; the resulting error on the main panel is smaller than point size. (c) The spatial profile $\delta \tilde \psi(r,k_\mathrm{d},\epsilon_\mathrm{p})$ is plotted for $k_\mathrm{d}=1$ (red) and $k_\mathrm{d}=3$ (blue) for resonant drive; the infinite-system expectation is shown as a dash-dotted line. Here $r_\mathrm{d}=5$, $T=150$, $N_r=200$, and $N_t=2\times 10^5$.}
\label{Fig3}
\end{figure}

To demonstrate a practical spectroscopic protocol for creating and detecting these core-bound states, we perform direct numerical simulations of the full GP equation for a finite-sized system with a realistic drive. We simulate Eq.~(\ref{GPE}), supplemented with a driving potential $U(r,z,t) = U_\mathrm{d}\exp(-r^2/r_\mathrm{d}^2)\sin(k_\mathrm{d}z)\sin(\epsilon_\mathrm{d} t)$, using a split-step Fourier-Bessel method; $r_\mathrm{d}$ is the drive radius and the drive amplitude $U_\mathrm{d}$ is chosen to remain in the linear-response regime (see Supplemental Material~\cite{supp}). The boundary conditions are periodic in $z$ and fixed in $r$, $\Psi(r=R)=0$.

To quantify the response to this drive, we compute the injected energy $E_\mathrm{i} \equiv E(T)-E(0)$ after a drive duration $T$, where $E(t)=\int \mathrm dV~\Psi^*(-\nabla^2+U+|\Psi|^2/2)\Psi$. In Fig.~\ref{Fig3}(a), we plot $E_\mathrm{i}$ versus $\Delta \epsilon_\mathrm{d}\equiv\epsilon_\mathrm{B}(k_\mathrm{d})-\epsilon_\mathrm{d}$ with (solid) and without (dashed) a vortex for $k_\mathrm{d}=1$ (red) and $k_\mathrm{d}=3$ (blue). In the presence of a vortex, there are two visible features: a narrow peak and a broad plateau. We identify the narrow peak as the lowest-lying varicose resonance (whose binding energy corresponds to the peak response $\Delta \epsilon_\mathrm{p}>0$) and the plateau as the scattered states (see End Matter Fig.~\ref{Fig4}). In the absence of a vortex, only the broad plateau remains.

In Fig.~\ref{Fig3}(b), we show the dispersion relation $\Delta\epsilon_\mathrm{p}(k_\mathrm{d})$ extracted from our numerical spectroscopy (black points), which is in excellent agreement with the linear-analysis calculation (dash-dotted line). This is promising for future experiments: within a GP description, nonlinear effects are indeed negligible when the drive is sufficiently weak. Finally, in Fig.~\ref{Fig3}(c) we directly verify the spatial nature of the excitation by plotting the spatial profiles $\delta \tilde\psi(r,k,\epsilon) \equiv \int \mathrm dt\int \mathrm dz~\sin(kz)\sin(\epsilon t+\delta)\Psi(r,z,t)$, where the phase shift $\delta$ is adjusted to maximize the signal. As expected, the profile is less localized for small $k_\mathrm{d}$ and approaches the $k=\infty$ linearized profile (dash-dotted line) for large $k_\mathrm{d}$.

In conclusion, we have studied two new families of core-bound waves on a GP vortex---the varicose and fluting modes---and established that each possesses an infinite sequence of geometrically-spaced bound states just below the Bogoliubov dispersion relation. Because these excitations are sensitive to vortex-core structure, they may provide a spectroscopic probe of microscopic physics in more complex systems, such as fermionic vortices~\cite{caroli1964bound}. An important next step would be to determine their lifetimes and interactions. Since dissipation in quantum turbulence is commonly framed in terms of interactions between Kelvin waves and phonons, it will be interesting to assess how varicose and fluting modes enter and possibly modify this dissipation pathway.

We thank Frédéric Chevy, Grisha Falkovich, and the members of the Ultracold Quantum Matter group at Yale University for discussions, as well as Vishvesha Sridhar and Siddhant Bhargava for contributions at early stages of this project. We thank Jack Harris for comments on the manuscript. This work was supported by the NSF (Grant No. PHY-2110303), the AFOSR (Grant No. FA9550-23-1-0605), and the David and Lucile Packard Foundation.



%


\newpage
\section{End matter}

\textbf{Appendix A.} Here, we provide an intuitive explanation for the effective potential $U_\mathrm{eff}(r)$. 
We rewrite Eq.~\eqref{u,v} as an equation for a spin-$1/2$ particle whose spin is coupled to a magnetic field $\mathbf B$:

\begin{equation}
    -\Delta k^2\begin{pmatrix}
        u\\v
    \end{pmatrix}= \bigg(-\frac{1}{r}\partial_r(r\partial_r) +V(r)-\mathbf B(r)\cdot \boldsymbol \sigma\bigg)\begin{pmatrix} u\\v\end{pmatrix}.
\end{equation}
Here $V(r)=(1+m^2)/r^2+2\psi_0^2-2+\sqrt{\epsilon^2+1}$, $\mathbf B(r) = \big(\psi_0^2(r),0,\epsilon-2m/r^2\big)$, $\Delta k^2\equiv k^2+1-\sqrt{\epsilon^2+1}$, and $\boldsymbol \sigma = (\sigma_1,\sigma_2,\sigma_3)$ are the Pauli matrices. At large $r$, the magnetic field angle $\theta\equiv \tan^{-1}(B_1/B_3)$ varies slowly $(\partial_r\theta\sim1/r^3$). This allows us to make the Born-Oppenheimer approximation that the spin aligns with the magnetic field, \emph{i.e.} $\mathbf B(r)\cdot \boldsymbol\sigma \approx |\mathbf B(r)|$. We then recover the effective potential $U_\mathrm{eff}(r)\equiv V(r)-|\mathbf B(r)|\sim-\alpha_m^2(\epsilon)/r^2$.

\begin{figure}[!hbt]
\includegraphics[width=\columnwidth]{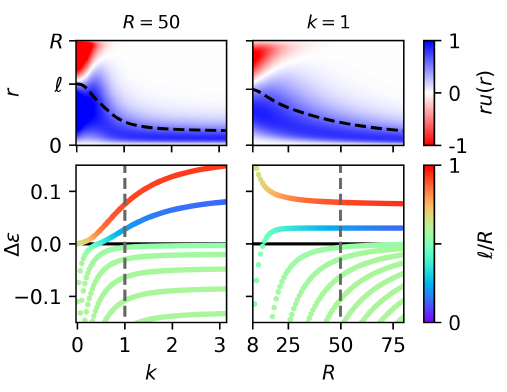}
\caption{\textbf{Finite-size effects on the \emph{m} $\boldsymbol{ ~=0}$ modes} (top) Varicose wave's spatial profile $u(r)$  and localization length $\ell$ (dashed line, see text) at fixed system size $R$ (left) and wavenumber $k$ (right). 
(bottom) Spectrum of excitations at fixed radius ($R=50$, left) and fixed wavenumber ($k=1$, right). The color indicates $\ell/R$.}
\label{Fig4}
\end{figure}
\textbf{Appendix B.} Here we discuss the fate of varicose waves in a finite system. To this end,
we diagonalize the Bessel-transformed Eq.~(\ref{u,v}), implementing the fixed boundary conditions $\psi_0(R)=u(R)=v(R)=0$. In Fig.~\ref{Fig4} (top), we show the lowest-lying varicose wave's spatial profile $u(r)$, versus $k$ at fixed $R$ (left) and versus $R$ at fixed $k$ (right). To capture the notion of localization, we introduce the parameter $\ell \equiv \int_0^Rr^2\psi_0|u-v|\mathrm{d}r/\int_0^Rr\psi_0|u-v|\mathrm{d}r$, which represents the typical radius of the excitation's density perturbation. We see that at low $k$ or small $R$, $\ell$ is of order $R$, \emph{i.e.} the excitation is delocalized.
For any given $k$, $\ell$ and $\Delta\epsilon$ become insensitive to the system size past a certain $R$.

In Fig.~\ref{Fig4} (bottom) we show the full $m=0$ spectrum for the finite-sized system. Phonons are delocalized, $\ell \sim R/2$ (green). As $R\to\infty$, they densely fill the region $\Delta\epsilon(k)<0$, whereas the varicose wave ($\ell\sim 1$, blue) approaches a system-size independent binding energy $\Delta\epsilon(k)>0$. 
Furthermore, the boundary condition at $r=R$ gives rise to an even lower-lying branch which is not vortex-specific and sits in the low-density region near the wall ($R-\ell\sim 1$, red)~\cite{Pikhitsa2019}. When this excitation overlaps with the varicose wave, the two hybridize. As Fig.~\ref{Fig4} (bottom) illustrates, this hybridization occurs at small $R$ (right), and small $k$ (left). For the finite-sized system, the varicose branch does not connect to the U(1) branch, and acquires a positive energy gap.

\begin{figure}[!hbt]
\includegraphics[width=1\columnwidth]{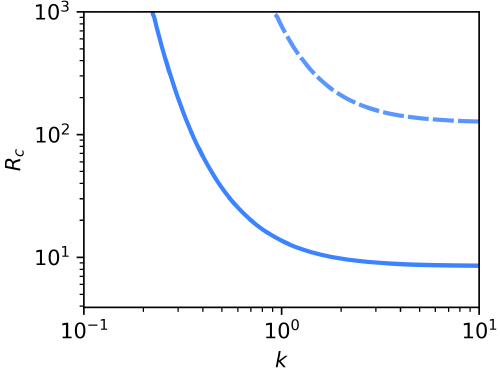}
\caption{\textbf{Finite-size effects on the varicose modes.} System radius $R_\mathrm{c}$ at which $\Delta\epsilon_0^{(0)} = 0$ (solid line) and $\Delta\epsilon_0^{(1)} = 0$ (dashed line) as a function of $k$.}
\label{Fig5}
\end{figure}

Fixing $k$, we may ask at what $R_\mathrm{c}$ this hybridization becomes significant, \emph{i.e.} $\Delta\epsilon=0$. For $R\gg R_\mathrm{c}$, the finite-size effects are small. In Fig.~\ref{Fig5}, we plot $R_\mathrm{c}$ for the lowest two varicose branches. For both branches, $R_\mathrm{c}$ asymptotes at a finite value as $k\rightarrow \infty$ and diverges as $k\rightarrow 0$.

\newpage
\widetext
\setcounter{equation}{0}
\setcounter{figure}{0}
\setcounter{table}{0}
\setcounter{page}{1}
\makeatletter
\renewcommand{\theequation}{S\arabic{equation}}
\renewcommand{\thefigure}{S\arabic{figure}}

\def\bibsection{\section*{\refname}} 
\pdfoutput=1

\makeatletter
\renewcommand \thesection{S\@arabic\c@section}
\renewcommand\thetable{S\@arabic\c@table}
\renewcommand \thefigure{S\@arabic\c@figure}
\renewcommand\theequation{S\@arabic\c@equation}
\renewcommand\theHfigure{S\@arabic\c@figure}
\renewcommand\theHequation{S\@arabic\c@equation}
\makeatother

\newpage

 \section{Supplemental Material}

\onecolumngrid

\section{I. Choice of Units}
The dimensionful time-dependent Gross-Pitaevskii equation
\begin{equation}
    i\hbar \partial_t \Phi(\mathbf r, t)=\left(-\frac{\hbar^2\nabla^2}{2M}+V_\mathrm{ext}(\mathbf r,t)+g|\Phi(\mathbf r,t)|^2\right)\Phi(\mathbf r, t)
\end{equation}
describes the behavior of a Bose-Einstein condensate wavefunction $\Phi$ in an external potential $V_\mathrm{ext}(\mathbf r,t)$ that may include both box trapping and forcing; $M$ is the boson mass, $N=\int |\Phi(\mathbf r, t)|^2\mathrm dV$ is the particle number, and $g$ is the coupling constant $4\pi \hbar^2 \as/M$ ($\as$ is the s-wave scattering length). For a cylindrical hard-wall potential (located at $r=R$), stationary solutions have the form $\Phi(\mathbf r,t)=e^{i(s\varphi - \mu t/\hbar)}\phi(r)$ \cite{Galilean}, where $\phi(r)$ solves the nonlinear eigenvalue problem
\begin{equation}
    \mu \phi(r)=\left(-\frac{\hbar^2}{2M}\left(\frac{1}{r}\partial_r (r\partial_r)-\frac{s^2}{r^2}\right) +g|\phi(r)|^2\right)\phi(r)\label{gGPE.S}
\end{equation}
with the boundary conditions that $\phi(r)$ is regular at $r=0$ and $\phi(R)=0$. For the infinite system, this second boundary condition becomes $\phi(r)\to 1$ as $r\to\infty$.

\begin{figure}[b!]
    \centering
    \includegraphics[width=.9\linewidth]{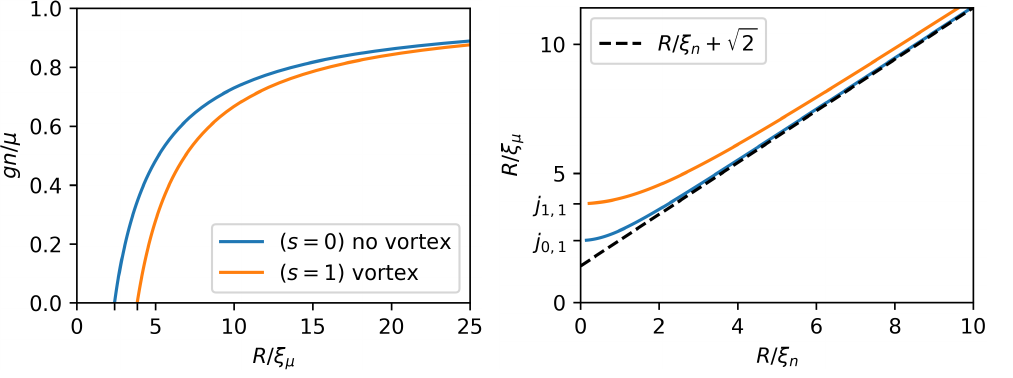}
    \caption{Choice of length scale unit. (Left) Energy scale ratio $gn/\mu$ as a function of the dimensionless boundary radius $R/\xi_{\mu}$.  (Right) Relation between the boundary in the parameter-dependent unit $R/\xi_{n}$ and solution-dependent unit $R/\xi_{\mu}$. The blue and orange colors correspond respectively to the no-vortex ($s=0$) and vortex ($s=1$) cases; $j_{i,j}$ is the $j^{\text{th}}$ zero of the $i^{\text{th}}$ Bessel function, $J_{i}$.}
    \label{fig:Supp0}
\end{figure}

We transform the time-independent GPE Eq.~(\ref{gGPE.S}) into the dimensionless form used in the text by defining a length scale $\xi_{\mu} = \sqrt{\frac{\hbar^2}{2M \mu}}$ and a dimensionless wavefunction $\psi_0(r/\xi_\mu)=\sqrt{\frac{g}{\mu}}\phi(r)$. This turns Eq.~\eqref{gGPE.S} into
\begin{align}
    \left(-\frac{1}{\tilde r}\partial_{\tilde{r} }(\tilde{r}\partial_{\tilde{r}})+\frac{s^2}{\tilde r^2}-1+|\psi_0(\tilde{r})|^2\right)\psi_0(\tilde{r})=0,\label{mGPE.S}
\end{align}
subject to the boundary conditions $\psi_0(0)=\psi_0(R/\xi_\mu)=0$ in the $s=1$ (vortex) case; for $s=0$ (no vortex), $\psi'_0(0)=\psi_0(R/\xi_\mu)=0$. Here we have introduced the dimensionless coordinate $\tilde r = r/\xi_\mu$. It is theoretically simpler to use $\xi_\mu$ rather than the experimentally convenient $\xi_{n}\equiv\hbar/\sqrt{2Mgn}$ (where $n\equiv N/V$ and $V$ is the box volume). The mapping between these two length scales is explicitly given by the normalization condition $gN/\mu = \int |\psi_0(r/\xi_\mu)|^2 \textrm{d}V$. Fig.~\ref{fig:Supp0} depicts this mapping; in the large-$R$ limit, $gn/\mu\rightarrow 1-\frac{2\sqrt{2}}{R/\xi_{\mu}}$ and $R/\xi_{\mu}\rightarrow R/\xi_{n}+\sqrt{2}$.

\section{II. Linear analysis for the infinite system}

\begin{figure}[b!]
\centering
\includegraphics[width=.9\linewidth]{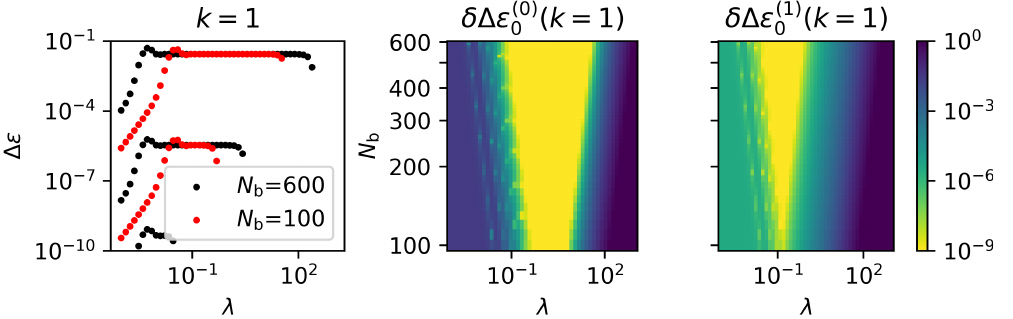}
\caption{\textbf{Finite Basis Scan} (left)  The calculated values of $\Delta\epsilon$ at $m=0$, $k=1$ as a function of $\lambda$ for $\Nb=600$ and $\Nb=100$.  We see plateaus form at $\Delta\epsilon\approx2.75\times 10^{-2}$ and $3.44\times 10^{-6}$.  
(center) The difference in the lowest eigenvalue for $m=0, k=1$ from the extracted plateau value, $\delta\Delta\epsilon_0^{(0)}(\lambda,\Nb) \equiv |\Delta\epsilon_0^{(0)}(\lambda,\Nb)-2.74849740\times 10^{-2}|$, varying basis size $\Nb$ and the length scale of the basis $\lambda$.  (right) The difference in the second-lowest eigenvalue from the extracted plateau value, $\delta\Delta\epsilon_0^{(1)}(\lambda,\Nb) = |\Delta\epsilon_0^{(1)}(\lambda,\Nb)-3.4442\times 10^{-6}|$.}
\label{fig:Supp1}
\end{figure}

To find bound states with a given $m$ for a vortex in an infinite system (Eq.~(3) in the main text), we project the solution space onto a finite-dimensional space (of dimension $2\Nb$) spanned by the following weighted Laguerre polynomials:  
\begin{align}
    \mathbf{e}^m_\nu = a_\nu^{2|m+1|}\lambda(\lambda r)^{|m+1|}e^{-\lambda r/2} L_{\nu}^{2|m+1|}(\lambda r)\begin{pmatrix}1\\0\end{pmatrix},\,\,\,\mathbf{e}^m_{\Nb+\nu} = a_\nu^{2|m-1|}\lambda(\lambda r)^{|m-1|}e^{-\lambda r/2} L_{\nu}^{2|m-1|}(\lambda r)\begin{pmatrix}0\\1\end{pmatrix},\label{eq:lbasis}
\end{align}
where $\nu\in \{0,1,...,\Nb-1\}$, $a_\nu^m=\sqrt{\frac{\nu!}{(\nu+m)!}}$, and $\lambda$ is a variational length scale. Our game plan is to find a regime of $\lambda$ where $\Delta\epsilon$ plateaus (with respect to $\lambda$), and extrapolate $\Nb$ to large values (see ~\cite{Alhaidari2008_2,Almarzoug2011_2}).

Defining an inner product $\braket{(u',v'),(u,v)} \equiv \int r (u'^*u+v'^*v)\mathrm{d}r$, the overlap matrix $\mathcal{S}^m_{ij} \equiv \braket{\mathbf{e}^m_i,\mathbf{e}^m_j}$ takes the block-diagonal form:
\begin{eqnarray}
    \mathcal{S}^m &=& \begin{pmatrix}S^{2|m+1|} &&0\\0&&S^{2|m-1|}
        \end{pmatrix}\\
    S^{m}_{\nu,\nu'} &=& (2\nu+m+1)\delta_{\nu,\nu'}-\sqrt{\nu(\nu+m+1)}\delta_{\nu,\nu'+1}-\sqrt{\nu'(\nu'+m+1)}\delta_{\nu+1,\nu'}
\end{eqnarray}
where $S^m$ is an $\Nb\times \Nb$ tridiagonal matrix.

The `potential' part of Eq.~(3),
\begin{align}
    \mathcal{U} \equiv \begin{pmatrix}
        2\psi_0(r)^2-1&&-\psi_0(r)^2\\-\psi_0(r)^2&&2\psi_0(r)^2-1
    \end{pmatrix},
\end{align}
has matrix elements that we calculate with a Gauss-Laguerre quadrature: we define the $\Nb\times\Nb$ matrix $P^m$, with matrix elements $P^m_{\nu,\nu'} \equiv\sum_{\mu}\Omega^m_{\nu,\mu} \omega^m_\mu\psi_0(\omega^m_\mu/\lambda)^2\Omega^m_{\nu', \mu}$, where the columns of $\Omega^m$ are the eigenvectors of $S^{m}$ and $\omega^m$ are the eigenvalues.  This form is an approximation of $\int_0^\infty r\psi_0(r)^2 \lambda^2 a^m_{\nu} a^m_{\nu'} (\lambda r)^{m} e^{-\lambda r} L_\nu^{m}(\lambda r) L_{\nu'}^{m}(\lambda r)\textrm{d}r$.

Practically, to calculate $\psi_0(r)^2$ at the quadrature points, $\omega^m/\lambda$, we numerically estimate $\psi_0(r)$ up to $r=16$, and use the asymptotic expansion, $\psi_0(r)\approx\psi_0^\mathrm{approx}(r)$ where $\psi_0^\mathrm{approx}(r)\equiv 1-\frac{1}{2r^2}-\frac{9}{8r^4}+...$ truncated to order $r^{-12}$ for $r>16$; the error is $e(r)\equiv |\psi_0^{\mathrm{approx}}(r)^2-\psi_0(r)^2| \lesssim 10^{-9}$ for all $r$, ensuring that the first-order correction in $e(r)$ to $\Delta\epsilon$ is $\lesssim 3\times10^{-9}$.

For $m<0$,  the transfer matrix $\mathbf{e}_{\nu
+\Nb}^m = \sigma_x \sum_{\nu'}T^m_{\nu,\nu'}\mathbf{e}_{\nu
'}^m$ can be analytically found via recurrence relations ($\sigma_x$ is the Pauli matrix).  For $m=0$ we set $T^0$ to the identity.

Thus, for $m\leq0$, $\mathcal{U}$ projected onto the $\{\mathbf{e}^m\}$ basis yields matrix elements $\mathcal{V}^{m}_{ij} = \braket{\mathbf{e}^m_i,\mathcal{U}\mathbf{e}^m_j}$, so that

\begin{align}
    \mathcal{V}^{m}&\approx\begin{pmatrix}
        2P^{2|m+1|}-S^{2|m+1|}&&-P^{2|m+1|} (T^m)^T\\-T^mP^{2|m+1|}&&2P^{2|m-1|}-S^{2|m-1|}
    \end{pmatrix}
\end{align}
Under the (time-reversal) symmetry of Eq.~(3), $u\leftrightarrow v$, $\epsilon\leftrightarrow-\epsilon$, and $m\leftrightarrow -m$, $m>0$ solutions are included within the  $m<0$ eigenvalue problems.  We impose the requirement $\int (|u|^2-|v|^2) \mathrm{d}V >0$ to get the sign of $m$.

The `kinetic' part of Eq.~(3)
\begin{align}
    \mathcal{H}^m_\mathrm{kin}&\equiv\begin{pmatrix}-\frac{1}{r}\partial_r(r\partial_r)+\frac{(1+m)^2}{r^2} &&0\\0&&-\frac{1}{r}\partial_r(r\partial_r)+\frac{(1-m)^2}{r^2}\end{pmatrix}
\end{align}
has matrix elements $\mathcal{K}^m_{ij}= \braket{\mathbf{e}^m_i,\mathcal{H}^m_\mathrm{kin} \mathbf{e}^m_j}$, so that
\begin{align}    
    \mathcal{K}^m &=\begin{pmatrix}K^{2|m+1|}_{\nu,\nu'} &&0\\0&&K^{2|m-1|}_{\nu,\nu'} \end{pmatrix},
\end{align}
and $K^m$ is an $\Nb\times\Nb$ tridiagonal matrix with coefficients
\begin{align}
    K^{m}_{\nu,\nu'} = \frac{\lambda^2}{4}(2\nu+m+1)\delta_{\nu,\nu'}+\frac{\lambda^2}{4}\sqrt{\nu(\nu+m+1)}\delta_{\nu,\nu'+1}+\frac{\lambda^2}{4}\sqrt{\nu'(\nu'+m+1)}\delta_{\nu+1,\nu'}.
\end{align}
To obtain $\epsilon$ at fixed $k$, we solve the generalized eigenvalue problem for the $2\Nb$-dimensional vector $\mathbf{w}$,
\begin{equation}
    (\mathcal{K}^m+\mathcal{V}^{m}+k^2\mathcal{S}^m)\mathbf{w} =  \epsilon\begin{pmatrix}S^{2|m+1|}&&0\\0&&-S^{2|m-1|}
    \end{pmatrix}\mathbf{w}.
\end{equation}

In Fig.~\ref{fig:Supp1}, we show the eigenvalue spectrum for $m=0$ at $k=1$, for various basis sizes $\Nb\in[100,600]$, and $\lambda \in[10^{-3},10^3]$.  We see plateaus for $\Delta\epsilon$ with respect to $\lambda$; the width of those plateaus gets larger with basis size, allowing the  extrapolation to the $\Nb\rightarrow\infty$ limit. We interpret these plateau values to be the lowest-lying and second-lowest-lying varicose modes' binding energies.  In the main text's Fig.~(2), we used $\Nb=800$.

\section{III. Numerical spectroscopy}

\begin{figure}[hbt!]
\includegraphics[width=0.8\columnwidth]{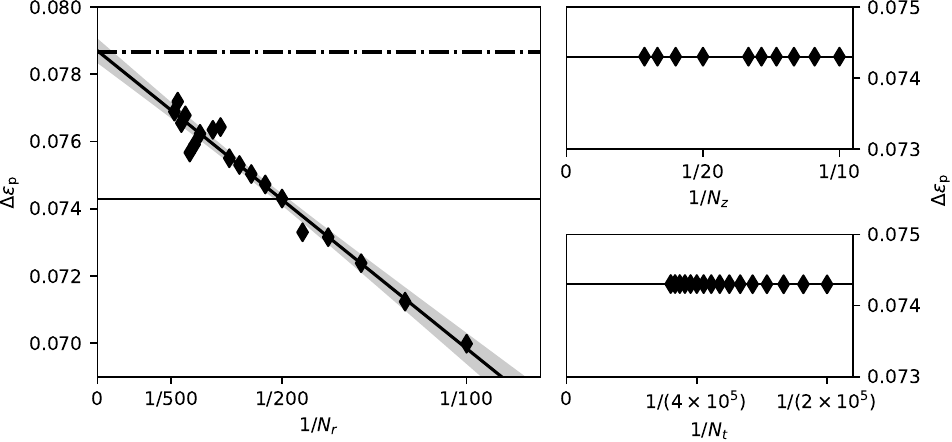}
\caption{\textbf{Discretization error.} (Left) varicose peak position $\Delta \epsilon_\mathrm{p}$ obtained from numerical spectroscopy versus $N_r$, with $N_z=12$ and $N_t=3.5\times10^5$; (top right) $\Delta \epsilon_\mathrm{p}$ versus $N_z$ with $N_r=200$ and $N_t=3.5\times 10^5$; (bottom right) $\Delta \epsilon_\mathrm{p}$ versus $N_t$ with $N_r=200$ and $N_z=12$. In all plots the horizontal line indicates $\Delta\epsilon_\mathrm{p}$ for $N_r=200$, $N_z=12$, and $N_t=3.5\times 10^5$. The linear fit in the left plot gives an $N_r\rightarrow \infty$ value of $\Delta \epsilon_\mathrm{p} = 0.0790\pm 7\times 10^{-4}$ (with error given by the $98\%$ confidence interval of the linear fit), in agreement with the prediction from the finite-system linear analysis of $0.0786$ (dash-dotted line). Throughout this figure, $T=80$ and $k_\mathrm{d}=3$.}
\label{FigS3.1}
\end{figure}

Here we provide additional details on the numerical spectroscopy scheme (see Fig.~3 of the main text). 

We use a split-step Fourier-Bessel method to simulate the dimensionless GP equation
\begin{equation}
    i\partial_t \Psi(r,\varphi,z,t) = \left(-\frac{1}{r}\partial_r(r\partial_r)-\frac{\partial^2_\varphi}{r^2}-\partial^2_z+U(r,z,t)+|\Psi(r,\varphi,z,t)|^2\right)\Psi(r,\varphi,z,t),
    \label{sim GPE}
\end{equation}
where $U(r,z,t)=U_\mathrm{d}\exp(-r^2/r_\mathrm{d}^2)\sin(k_\mathrm{d}z)\sin(\epsilon_\mathrm{d} t)$ is the potential to excite the varicose mode (we use $r_\mathrm{d}=5$ everywhere except in Fig.~\ref{FigS3.3}), and a cylindrical trapping potential is imposed by the boundary condition $\Psi(r=R=50)=0$.
The symmetries of Eq.~\eqref{sim GPE} allow for the problem to be simplified in two ways. First, we need only to consider Bessel functions of order $s$, where $s$ is the azimuthal wavenumber (recall that $s=1$ with a vortex and $0$ without), since azimuthal symmetry forbids coupling between Fourier-Bessel components with different $s$. Second, we are allowed to use a periodic boundary condition along $z$: $\Psi(z=0)=\Psi(z=2\pi/k_\mathrm{d}),\;\partial_z\Psi(z=0)=\partial_z\Psi(z=2\pi/k_\mathrm{d})$. This restricts the $z$ wavenumbers to be integer multiples of $k_\mathrm{d}$, which is permissible because of the discrete translation symmetry along $z$ of the potential $U$. This may be counterintuitive, since some physical processes involve higher momentum modes breaking into lower momentum ones (\emph{e.g.} Beliaev damping); given Eq.~\eqref{sim GPE} and an initial condition that is homogeneous along $z$, however, subharmonic generation does not occur in this classical model.

We discretize Eq.~(\ref{sim GPE}) on a grid of $N_r\times N_z$ points and $N_t$ time steps. We use a basis of $\lfloor3N_r/4\rfloor\times \lfloor 3N_z/4\rfloor$ Bessel-Fourier components to prevent aliasing. To estimate errors stemming from this discretization, we compute spectra with various values of $N_r$, $N_z$, and $N_t$. In Fig.~\ref{FigS3.1}, we plot the varicose peak position $\Delta\epsilon_\mathrm{p}$ against $1/N_r$ (left), $1/N_z$ (top right) and $1/N_t$ (bottom right). The effect of $N_r$ is the most significant. We empirically find that $\Delta\epsilon_\mathrm{p}$ follows a relation of the type $c_1/N_r+c_2$. We use a linear fit to extrapolate the peak position $c_2$ for $N_r\rightarrow \infty$. The right panels illustrate that $N_z$ and $N_t$ have a negligible impact on the peak position compared to $N_r$.

\begin{figure}[hbt!]
    \includegraphics[width=\columnwidth]{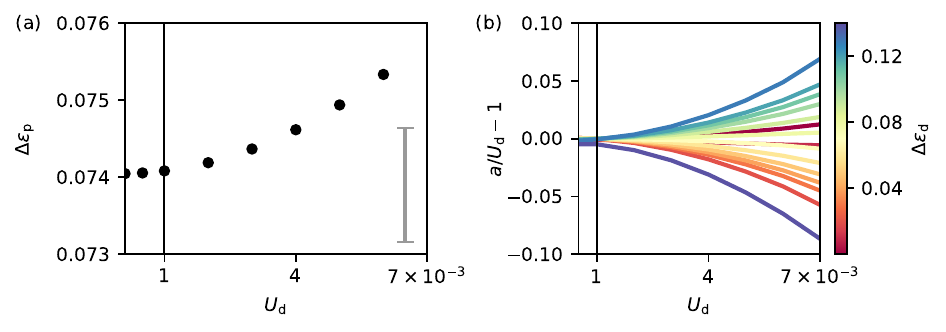}
    \caption{\textbf{Error arising from nonlinear response.} (a) Varicose peak position $\Delta\epsilon_\mathrm{p}$ versus drive amplitude $U_\mathrm{d}$; the gray error bar is the error due to the discretization of $N_r$ (see Fig.~\ref{FigS3.1}). (b) Nonlinear response $a/U_\mathrm{d}-1$, with $a(U_\mathrm{d})\equiv\sqrt{\Ei}\big/\left(\frac{\dd\sqrt{\Ei}}{\dd U_\mathrm{d}}\Big|_{U_\mathrm{d}=0}\right)$, as a function of $U_\mathrm{d}$ for various $\Delta \epsilon_\mathrm{d}$. In both plots, $k_\mathrm{d}=3$, $N_r=200$, $N_z=12$, $T=100$, and $N_t=2\times 10^5$; the vertical solid line indicates $U_\mathrm{d}=10^{-3}$ used in Fig.~3 in the main text (and elsewhere).}
    \label{FigS3.2}
\end{figure}

\begin{figure}[hbt!]
\includegraphics[width=0.8\columnwidth]{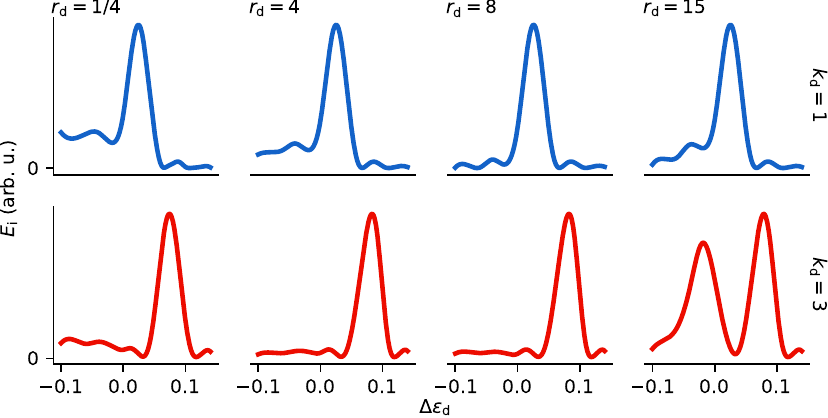}
\caption{\textbf{Influence of the drive radius $r_\mathrm{d}$ on the injected energy spectra.} Injected energy $\Ei$ versus $\Delta\epsilon_\mathrm{d}$ for drive radii $r_\mathrm{d}=1/4,4,8,$ and $15$ (increasing from left to right). The top row shows $k_\mathrm{d}=1$ (blue) and the bottom shows $k_\mathrm{d}=3$ (red). For these spectra, $T=150$ (the Fourier-limited FWHM is $\approx 0.04$), $N_t=2\times 10^5$, $N_r=200$, and $N_z=12$.}
\label{FigS3.3}
\end{figure}

In Fig.~\ref{FigS3.2}, we estimate the effects of the weak nonlinear response on our spectroscopy. First, we perform the simulations for various $U_\mathrm{d}$ and extract the corresponding varicose peak position $\Delta\epsilon_\mathrm{p}$ [Fig.~\ref{FigS3.2}(b)].  This shows that nonlinear effects are negligible provided $U_\mathrm{d}$ is small enough; for reference, the error due to $N_r$ is shown as the gray error bar. In Fig.~3 of the main text, we use $U_\mathrm{d}=10^{-3}$ (solid vertical line). 

More generally, to quantify nonlinearity, we define $a(U_\mathrm{d})\equiv\sqrt{\Ei}\big/\left(\frac{\dd\sqrt{\Ei}}{\dd U_\mathrm{d}}\Big|_{U_\mathrm{d}=0}\right)$. This is a measure of the amplitude of the response, normalized so that $\frac{\dd a}{\dd U_\mathrm{d}}\Big|_{U_\mathrm{d}=0}=1$. In Fig.~\ref{FigS3.2}(b), we plot the nonlinear part $a/U_\mathrm{d}-1$ versus $U_\mathrm{d}$ for a given spectrum (here $\epsilon_\mathrm{B}-\Delta\epsilon_\mathrm{d}$ is the drive frequency).

Finally, we investigate how the spectrum depends on the drive radius $r_\mathrm{d}$, since it affects the spatial overlap of the drive with the different modes of the system. In Fig.~\ref{FigS3.3}, we show several spectra for $k_\mathrm{d}=1$ (top row, blue) and $k_\mathrm{d}=3$ (bottom row, red) for various $r_\mathrm{d}$.  
For small $r_\mathrm{d}$, the drive couples well to many modes. We see this on the leftmost plots of Fig.~\ref{FigS3.3}: for both $k_\mathrm{d}=1$ and $k_\mathrm{d}=3$, the phonon plateau is very visible. As we increase $r_\mathrm{d}$, the phonon ridge begins to die down, and the varicose mode becomes more prominent. For excessively large $r_\mathrm{d}$, the drive couples well to the lower-lying phonons. This is especially striking in the bottom right plot ($k_\mathrm{d}=3$, $r_\mathrm{d}=15$), where the lowest phonon modes come together to form a second peak.

It is interesting to contrast the dependence on $r_\mathrm{d}$ between $k_\mathrm{d}=1$ and $k_\mathrm{d}=3$. For $k_\mathrm{d}=1$, the varicose mode is most prominent when $r_\mathrm{d}=8$. For $k_\mathrm{d}=3$, however, the $r_\mathrm{d}=4$ and $r_\mathrm{d}=8$ curves are similar. At $r_\mathrm{d}=15$, the $k_\mathrm{d}=3$ varicose peak is almost overshadowed by the low-energy phonon peak, whereas the $k=1$ peak is still prominent. This indicates that larger drive radii couple better to longer-wavelength varicose waves, reflecting the varicose mode's progressive delocalization at small $k_\mathrm{d}$.

\end{document}